\documentclass[conference]{ieeeconf}  %

\IEEEoverridecommandlockouts                              %

\usepackage{graphics} %
\usepackage{times} %
\usepackage{amsmath, amssymb} %
\usepackage{booktabs} %
\usepackage[group-digits=false]{siunitx} %
\usepackage{graphicx} %
\usepackage{hyperref} 
\usepackage[capitalise]{cleveref}
\usepackage{algorithm}
\usepackage{algpseudocode}
\usepackage{mathtools}
\usepackage{color, soul}

\newtheorem{definition}{Definition}
\newcommand{\todo}[1]{\textcolor[rgb]{0.6,0.2,0.2}{{{#1}}}}

\newrobustcmd*{\bftabnum}{%
  \bfseries
  \sisetup{output-decimal-marker={\textmd{.}}}%
}
\title{\LARGE \bf
Polyhedral Collision Detection via Vertex Enumeration
}

\author{Andrew Cinar$^{1}$, Yue Zhao$^{2}$ and Forrest Laine$^{2}$%
\thanks{*This work was not supported by any organization}%
\thanks{$^{1}$Andrew Cinar is with the Department of Mechanical Engineering, Vanderbilt University,
        Nashville, TN 37235, USA
        {\tt\small al.cinar@vanderbilt.edu}}%
\thanks{$^{2}$Yue Zhao and Forrest Laine are with the Department of Computer Science, Vanderbilt University,
        Nashville, TN 37235, USA
        {\tt\small \{forrest.laine, yue.zhao\}@vanderbilt.edu}}%
}

\begin{document}

\mbox{\begin{minipage}[c]{\textwidth}
 \begin{center}
© 2025 IEEE.  Personal use of this material is permitted.  Permission from IEEE must be obtained for all other uses, in any current or future media, including reprinting/republishing this material for advertising or promotional purposes, creating new collective works, for resale or redistribution to servers or lists, or reuse of any copyrighted component of this work in other works. \\
\hfill \break
Accepted version.
 \end{center}
\end{minipage}
}

\maketitle
\thispagestyle{empty}
\pagestyle{empty}
\begin{abstract}
Collision detection is a critical functionality for robotics. The degree to which objects collide cannot be represented as a continuously differentiable function for any shapes other than spheres.
This paper proposes a framework for handling collision detection between polyhedral shapes. We frame the signed distance between two polyhedral bodies as the optimal value of a convex optimization, and consider constraining the signed distance in a bilevel optimization problem. To avoid relying on specialized bilevel solvers, our method exploits the fact that the signed distance is the minimal point of a convex region related to the two bodies. Our method enumerates the values obtained at all extreme points of this region and lists them as constraints in the higher-level problem. We compare our formulation to existing methods in terms of reliability and speed when solved using the same mixed complementarity problem solver. 
We demonstrate that our approach more reliably solves difficult collision detection problems with multiple obstacles than other methods, and is faster than existing methods in some cases.

\end{abstract}

\section{INTRODUCTION}

Handling collision detection in trajectory planning is hard, primarily because the degree to which two objects intersect in physical space (the signed distance) generally can not be represented by a continuous and differentiable function, as is typically required by solvers of constrained optimization problems.  
To the best of our knowledge, only the distance between spheres can be expressed in this way.
For example, two circles in the plane with centers $c_1$ and $c_2$, and radii $r_1$ and $r_2$ are intersecting if and only if $\|c_1 - c_2\| \le r_1+r_2$. The situation is not as simple if the circles are replaced with a triangle and a rectangle. 
The most common way to circumvent this issue is by approximating exact collision detection, representing the relevant geometries as union of spheres \cite{spica_real-time_2020, breton_advances_2011, reiter_hierarchical_2022, nair_stochastic_2022, van_den_berg_reciprocal_2011, fan_collision_2021, li_cooperative_2023, ren_research_2019, chao_design_2018}. 
This approach can lead to excessive conservatism for poor approximations, or accurately modeling the geometry can be %
expensive.
 
Several approaches use optimization-based set intersection to formulate the collision detection \cite{zhang_optimization-based_2021, schulman_motion_2014}, which allows treating obstacles as constraints. 
When two bodies are represented via convex sets, notions of distance or intersection between the sets is the solution to a convex optimization problem. 
While the solutions to these convex optimization problems are generally not analytic expressions, they can be embedded into higher-level optimization problems, giving rise to a bilevel optimization problem. 
Most methods taking the bilevel approach either, (i) encode the entirety of the Karush-Kuhn-Tucker (KKT) conditions for the low-level problem as constraints in the high-level problem, or (ii) wrap the low-level convex optimization problem in a functional form which will \emph{appear} to be continuous and differentiable to the solver.
Both approaches come with significant challenges:

(i) The KKT conditions of the low-level problem include complementarity conditions that violate the constraint requirements for interior point solvers such as IPOPT \cite{wachter_implementation_2006}.
Although complementarity conditions may not violate some of the most general constraint qualifications like Guignard, most solvers require additional assumptions on the type of constraints allowed, and have difficulty with complementarity constraints. This often results in suboptimal solutions to the higher-level problem, or even failure to find a solution.

(ii) Attempting to wrap the low-level optimization collision detection problem as a continuous and differentiable function also faces significant difficulty, because most approaches invoke the implicit function theorem \cite{zimmermann_differentiable_2022, le_cleach_single-level_2023, tracy2023differentiable, lee_uncertain_2023} to produce derivatives of the low-level optimal decision variables with respect to the high-level decision variables, while high-level decision variables appear as problem data (parameters) in the low-level. 
This approach is fundamentally flawed, because when inequality constraints are \emph{weakly active} at the solution of the low-level problem, the implicit function theorem cannot be invoked. 
Derivatives of the optimal decision variables and objective value with respect to the problem data at these points generally do not exist. 
The set of points in problem data space where this occurs is normally of measure zero, so it is tempting to ignore any issues of non-differentiability. 
However, the points at which a low-level solution will not be differentiable are frequently encountered in bilevel collision detection.

We enumerate all vertices of the feasible region of a linear program whose solution is the signed distance between two polyhedra. We propose supplementing the low-level collision detection problem by introducing slots that dynamically incorporate constraints corresponding to multiple vertices (not only optimal vertices) of the feasible region of the  problem, which are ranked based on the proximity to the optimal vertex. %
We compare our vertex enumeration formulation  against various baseline formulations using a standard off-the-shelf mixed complementarity problem solver. We discover that our approach offers multiple benefits and achieves collision-free trajectories in very challenging settings.

In this paper, our explicit contributions are the following:
\begin{itemize}
    \item %
    A procedure and code repository\footnote{The code is available at \href{https://github.com/VAMPIR-Lab/PolyPlanning}{https://github.com/VAMPIR-Lab/PolyPlanning}.} for replacing nonsmooth constraints with multiple smooth constraints by enumerating the values obtained at all vertices of the feasible region of signed distance between polyhedra.
    \item Demonstrations of our procedure on several challenging settings and direct comparisons with existing baseline formulations in terms of reliability and performance.
\end{itemize}

\section{PROBLEM DEFINITION}

Our motivation is solving an optimization problem with constraints for enforces collision avoidance between polyhedral objects. This section outlines our high-level optimization problem and collision detection formulation.
\todo{}

\subsection{Problem Formulation}
For simplicity, we focus on discrete-time, deterministic trajectory optimization with nonlinear discrete dynamics. Let dynamics of the ego object (or just ego) be,
\begin{equation}
\label{eq:dynamics}
\begin{aligned}
x_{t} = f(x_{t-1}, u_{t-1}) ,\quad \forall t\in\{1,\dots, T\}, 
\end{aligned}
\end{equation}
representing second-order rigid-body dynamics with full degrees of freedom (DoF), where $x_t \in \mathbb{R}^{6 (d - 1)}$ is the state at time $t$ in $d\in\{2, 3\}$-dimensional space with full-DoF second-order controls $u_t \in\mathbb{R}^{3 (d - 1)}$. The initial state and control of the ego is $x_0$ and $u_0$.
We use full state and control parametrization for nonlinear optimization, the decision variable is defined as $z_t  = \begin{bmatrix}x_t& u_t\end{bmatrix}^\intercal$ for all $t \in\{1,\dots, T\}$. The complete decision variable $z = \{z_1, \dots, z_T\}\in\mathbb{R}^n$ represents the whole trajectory, where $n = T \times 9 (d - 1)$. This decision variable is subject to \cref{eq:dynamics}, and other constraints such as saturation of controls, and collision constraints. 

We introduce a signed distance 
function $\mathrm{sd}(x_t)$ to detect collisions between two polyhedral objects, which we will define later in this section. We focus on single ego and obstacle here, represented by one polyhedral object each, but this formulation can easily be extended to multiple egos and obstacles. The trajectory optimization problem can be formulated as follows:
\begin{equation} \label{opt}
    \begin{aligned} 
        \min_{z\in\mathbb{R}^n} F(z) \quad
        \text{s.t.} \quad 0 &= x_{t}-f(x_{t-1}, u_{t-1}), \\
        \quad 0 &\leq g_t(x_t, u_t), \\
        \quad 0 &\leq \mathrm{sd}(x_t) , %
        \quad  \quad \forall t \in\{1,\dots, T\}.
    \end{aligned}
\end{equation}
The cost function of the high-level problem is $F(z)$.
If the ego or the obstacle is represented as a union of multiple polyhedral shapes, the collision detection constraint $\mathrm{sd}(x_t) \ge 0$ should be applied to each pair of polyhedral shape from the ego and the obstacle. This union does not need to be convex, however as more shapes are added the number of pairs grow polynomially. 

\subsection{Convex Object Processing}

Consider a convex object $\mathcal{O}$ described by $m$ continuous and concave functions $h_{j} : \mathbb{R}^d \to \mathbb{R}$:
\begin{equation} \label{eq:obj}
    \mathcal{O} \coloneqq \left\{ 
    \begin{aligned}
        p\in \mathbb{R}^d : h_j(p) \ge 0, \quad \forall j \in \{1,...,m\} 
    \end{aligned}
    \right\} ,
\end{equation}
Specifically, we require these concave functions to be affine or have a maximum.
Given a state vector $x_t$ including position and orientation, we get a translation vector  $l(x_t) \in \mathbb{R}^d$ and a rotation matrix $R(x_t) \in \mathbb{R}^{d\times d}$ that rigidly transforms $\mathcal{O}$ into $\mathcal{O}^{x_t}$, letting  ${h_j^{ x_t}(p)\coloneqq h_j \left( R(x_t)^\intercal(p-l(x_t)) \right)}$ :
\begin{equation} \label{eq:obj_xt}
    \mathcal{O}^{x_t} \coloneqq \left\{ 
    \begin{aligned}
        p\in \mathbb{R}^d : h_j^{x_t}(p) \ge 0, \quad \forall j \in \{1,...,m\} 
    \end{aligned}
    \right\} ,
\end{equation}
Let $c_0 \in \mathbb{R}^d$ be an interior point of $\mathcal{O}$, and $c^{x_t}\coloneqq R(x_t)c_0+l(x_t)$ be the corresponding point after rigid transformation,  %
we can define a scaled object (parametrized closed convex set) with a parameter $\alpha=[-1, \infty)$:
\begin{align} 
\label{eq:obj_xt_alpha}
    \mathcal{O}^{x_t}(\alpha) &\coloneqq \left\{ 
    \begin{aligned}
        p\in \mathbb{R}^d : h_j^{x_t}(p; \alpha, p_j^*)  \ge 0, \\
        \forall j \in \{1,...,m\} 
    \end{aligned}
    \right\} ,\\
    \label{eq:para_convex_set2}
    h_j^{x_t}(p; \alpha, p_j^*) &\coloneqq h_j^{x_t}\left(p+\alpha(c^{x_t} -p_j^*)\right) + \alpha h_j^{x_t}(p_j^*).
\end{align}
where $p^*_j$ can be any feasible point when $h_j^{x_t}(p) \text{ is affine}$, because \cref{eq:para_convex_set2} simplifies to $h_j^{x_t}(p)+\alpha h_j^{x_t}(c) \ge 0$.  Otherwise $p^*_j={\mathrm{argmax}_p} \; h_j^{x_t}(p)$, solution to which always exists and unique.  Note when $\alpha=-1$, $p=c$ uniquely solves \cref{eq:para_convex_set2}  . 

Intuitively, the $\alpha$ is a scaling factor that  uniformly inflates (or deflates) the convex object. The object inflates with around the point $c$ when $\alpha>0$,  covering the entire space as $\alpha \to \infty$.  When $\alpha=0$, \cref{eq:obj_xt_alpha} is the same as  \cref{eq:obj_xt}. When $\alpha<0$, the object deflates to a singleton $\{c\}$ at $\alpha=-1$.  

\subsection{Signed Distance}\label{subsec:sd}
In our approach, similar to some others in the literature \cite{zhang_optimization-based_2021, le_cleach_single-level_2023, montaut_differentiable_2022, guthrie_differentiable_2022}, we think of the signed distance $\mathrm{sd}(x_t)$ as solving a parameterized convex optimization problem. 

\vspace{1em}
\begin{definition}[Signed Distance]
    \begin{equation} \label{eq:sd}
       \begin{aligned}
       \mathrm{sd}(x_t) := \min_{\alpha \in \mathbb{R}} \quad  \alpha \quad
            \mathrm{s.t.} \quad p \in \mathcal{O}_1^{x_t}(\alpha) \cap \mathcal{O}_2(\alpha).
        \end{aligned}
    \end{equation}
\end{definition} 
$\mathcal{O}_1^{x_t}(\alpha)$ is the translated and rotated ego parameterized by state vector $x_t$, $\mathcal{O}_2(\alpha)$ is the stationary obstacle, as defined in \cref{eq:obj_xt_alpha}. We assume the obstacle is stationary for simplicity sake, but the extension to moving objects is straightforward.  The value of the signed distance is found by solving the convex optimization problem represented by \cref{eq:sd}. When $\mathrm{sd}(x_t) > 0$, we inflate both objects to obtain an intersection point, indicating that they have not collided; when $\mathrm{sd}(x_t)=0$, they are just touching; when $\mathrm{sd}(x_t)<0$, they have penetrated each other.
For clarity, we focus on the case when every function $h_{j}$ is affine for both the ego and obstacle, we can derive the following from \cref{eq:obj_xt_alpha}:
\begin{equation} \label{eq:obj_xt_alpha_affine}
    \mathcal{O}_i^{x_t}(\alpha) \coloneqq \left\{ 
    \begin{aligned}
        &p\in \mathbb{R}^d : \\  &A_i^{x_t}p + b_i^{x_t}+\alpha(A_i^{x_t}c_i^{x_t} + b_i^{x_t}) \ge 0\} 
    \end{aligned}
    \right\} ,
\end{equation}
 $A_i^{x_t} \in \mathbb{R}^{m_i \times d}$, $b_i^{x_t} \in \mathbb{R}^{m_i}$, $c_i^{x_t} \in \mathbb{R}^d$, where  $m_i$ is the number of constraints for object $i$, in this case $i \in \{1,2\}$. Note that in the case of ego, $\mathcal{O}_1^{x_t}(\alpha)$, $A_1^{x_t}, b_1^{x_t}$ and $c_1^{x_t}$ are parameterized by $x_t$, but not for the obstacle, so $x_t$ is omitted from $\mathcal{O}_2(\alpha)$. Thus the corresponding problem is a parametric linear program (LP), derived from \cref{eq:sd} :

\begin{equation}
\begin{aligned}
\label{eq:sd_lp}
\mathrm{sd}(x_t) = \underset{z}{\min}\quad& \overbrace{\begin{bmatrix}0 &\dots & 0 & 1 
\end{bmatrix}}^{q_\mathrm{sd}^\intercal} w \\
\text{s.t.} \quad& \underbrace{\begin{bmatrix}A_1^{x_t} & A_1^{x_t} c_1^{x_t} + b_1^{x_t} \\ A_2 & A_2 c_2 + b_2 \end{bmatrix}}_{A_\mathrm{sd}\in\mathbb{R}^{m_\mathrm{sd}\times (d+1)}} w + \underbrace{\begin{bmatrix}b_1^{x_t} \\  b_2 \end{bmatrix}}_{b_\mathrm{sd}\in\mathbb{R}^{m_\mathrm{sd}}} \geq 0
\end{aligned}
\end{equation}
where $w=\begin{bmatrix} p & \alpha \end{bmatrix}^\intercal \in\mathbb{R}^{d+1}$ is the intersection point and the scaling factor, $m_\mathrm{sd}=m_1+m_2$. 
The signed distance function with this LP form is continuous but not differentiable everywhere. 

\textbf{Continuity:}
Note that the feasible region of \cref{eq:sd_lp} is a $d+1$, or $\{3, 4\}$-dimensional polyhedral region parameterized by $x_t$, and the minimum value of the objective function can always be attained at one or more vertices, called the optimal vertices, of this polyhedra. 
While the intersection point $p$ is not necessarily unique, the minimizer of $\alpha$ always exists and is continuous with respect to the problem parameters, ensuring that $\mathrm{sd}(x_t)$ is a well-defined continuous function.

\textbf{Non-differentiability:}
For a $\{3, 4\}$-dimensional polyhedron defined by $m_\mathrm{sd}$ inequality constraints, a vertex $w$ requires at least $\{3, 4\}$ inequalities to become equalities. Every vertex can be attained by solving a linear system $A w + b = 0$, where $A \in \mathbb{R}^{(d+1) \times (d+1)}$, $b \in \mathbb{R}^{d+1}$. $A$ and $b$ depend on which inequalities we select to become equalities. If we traverse all possible selections, we can get all vertices.

\textbf{Assignments:} We refer to one selection of $d+1$ constraints as an \emph{assignment}. There are a total of $\left(\substack{m_\mathrm{sd} \\ d+1}\right)$ assignments. Every assignment can uniquely determine the linear system above, and we refer to its solution $w=A^{-1}b$ as an \emph{assignment point}. But an assignment point may violate constraints that this assignment does not select, and we call this assignment \emph{infeasible}. Infeasible assignment points fall outside the feasible polyhedral region of \cref{eq:sd_lp}. 
Feasible assignment points must be vertices of the feasible region, some of which are optimal vertices, and the corresponding assignments  are optimal.

Given $x_t$, if there are only $d+1$ active constraints, there is only one optimal assignment, and we can get a unique assignment point $w$, so $\alpha$ is also unique, and the signed distance function $sd(x_t)$ is smooth.
However, when there are more than ${d+1}$ active constraints, the optimal assignment is not unique,
and $sd(x_t)$ is nonsmooth, because  although $\alpha$ corresponding to these optimal assignments have the same value,  they have different expressions and derivatives.

\begin{figure}[t]
\centering
\includegraphics[width=\linewidth]{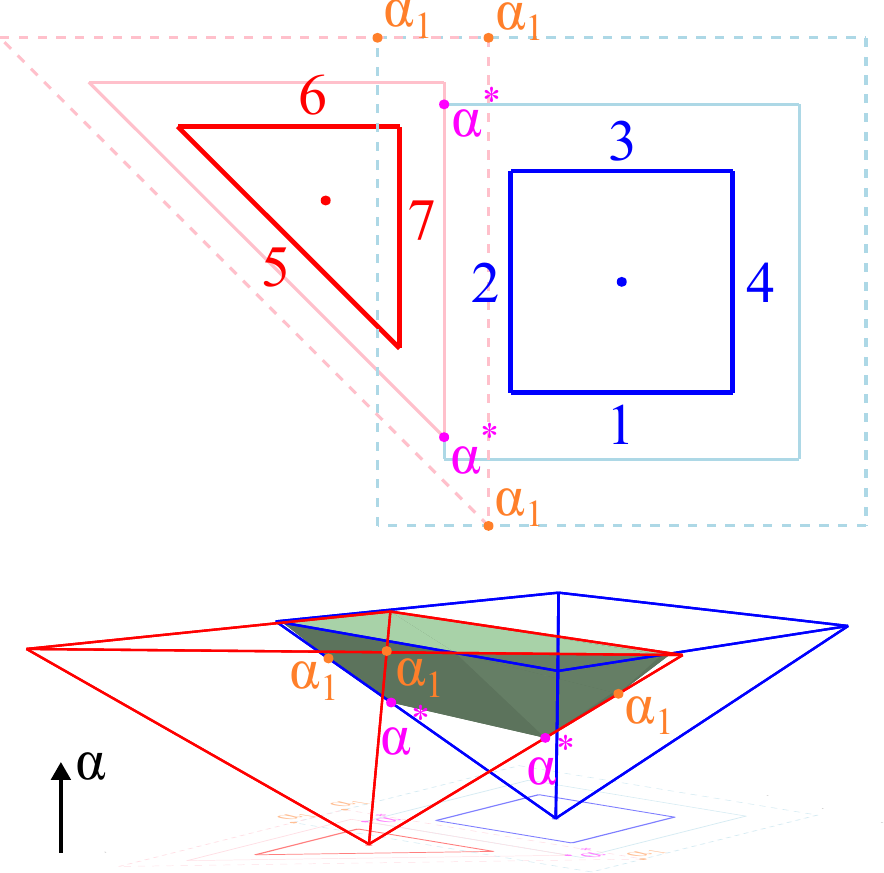}
\caption{ 
 On the top subfigure, the edges of the blue rectangle and the red triangle are numbered from $1$ to $7$. The solid faded lines show $\alpha=\alpha^*$ corresponding to optimal assignments $(\textcolor[rgb]{0,0,1}{2,3},\textcolor[rgb]{1,0,0}{7})$, $(\textcolor[rgb]{0,0,1}{2},\textcolor[rgb]{1,0,0}{5,7})$. The dashed lines show $\alpha=\alpha_1$  corresponding to assignments $(\textcolor[rgb]{0,0,1}{2,3},\textcolor[rgb]{1,0,0}{7})$, $(\textcolor[rgb]{0,0,1}{1},\textcolor[rgb]{1,0,0}{5,7})$, $(\textcolor[rgb]{0,0,1}{3},\textcolor[rgb]{1,0,0}{6, 7})$.  
On the bottom subfigure, the vertical axis represents $\alpha$. The polyhedral regions defined by the constraints of the rectangle and the triangle in this representation take the form of an upside down quadrilateral pyramid and triangular pyramid, respectively, and are unbounded in the positive direction. The light green region is their intersection, i.e., the feasible region of \cref{eq:sd_lp}.}%
\label{fig:geometry}
\end{figure}

In \cref{fig:geometry}, we show an example with two polygons and the corresponding 3-dimensional polyhedral LP space. There exist five feasible assignments, but only two values of $\alpha \in \{\alpha*, \alpha_1\}$. Clearly, $\alpha^*$ is the minimizer of \cref{eq:sd_lp}, i.e. the actual signed distance. When $\alpha=\alpha^*$, the intersection point $p$ can be any point on the segment  
corresponding to an edge in the 3-dimensional space. Any point on this edge is a solution to the LP. If the ego rotates slightly, this edge will degenerate to either one of the vertices that correspond to $\alpha=\alpha^*$. 

\section{Collision Detection via Vertex Enumeration}
\label{sec:method}

In the previous section we showed that collision detection related to two polyhedral objects is the problem of finding the minimal point of a polyhedral region. From this perspective, we can identify two contributing factors to the nonsmooth characteristic of the signed distance value, complicating the solution of \cref{opt}. The first stem is from the fact that the minimization of \cref{eq:sd_lp} is a linear program, and hence $\mathrm{sd}(x_t)$ can be equivalently expressed as minimization over the value obtained by the vertices of its feasible region. Although all vertices are smooth functions of the polyhedral objects, the min over vertices is nonsmooth. We address  this issue by simply listing the value at all (or most) vertices as constraints. This also has the desirable advantage of introducing nonlocal information to the solver, as discussed below. 
The other contributing factor to nonsmoothness is that the set of vertices is not stable, the number of vertices and which vertices are in this set may change.
Our method uses heuristic approaches to alleviate this issue. We present more details in the remainder of this section.

We address the nonsmooth minimizer over vertices by indexing the values obtained at all vertices of the feasible region of \cref{eq:sd_lp}, and including the value at these vertices (denoted $\mathrm{sd}_k(x_t)$ for the $k$-th vertex) as additional constraints to \cref{opt}.
For a region with $K$ vertices, we note that $\mathrm{sd}(x_t) = \min\{\mathrm{sd}_1(x_t),\dots,\mathrm{sd}_K(x_t)\} \geq 0$ is equivalent to  $\mathrm{sd}_k(x_t) \geq 0$  for all $k$. 

For two polyhedral bodies which always result in $K$ vertices of the feasible region of \cref{eq:sd_lp}, \cref{opt} becomes,
\begin{equation} \label{opt2}
    \begin{aligned} 
        \min_{z\in\mathbb{R}^n} F(z) \quad
        \text{s.t.} \quad 0 &= x_{t}-f(x_{t-1}, u_{t-1}), \\
        \quad 0 &\leq g_t(x_t, u_t), \\
        \quad 0 &\leq \mathrm{sd}_k(x_t), \quad \forall k \in \{ 1, \dots, K\},\\
        \quad  \forall t &\in\{1,\dots, T\}.
    \end{aligned}
\end{equation}
However, as the configuration of the polyhedral objects change, the number of vertices itself is clearly not constant, and $K$ is a function of $x_t$. This is a problem for optimization solvers that they do not support variable number of constraints. It is tempting to  constrain the value at all \emph{assignment points} (as defined in \cref{subsec:sd}) instead of strict vertices, since the number of all assignment points is independent of $x_t$, but not all assignment points are feasible.

To overcome this, we propose \emph{slots}, placeholders which are dynamically filled with appropriate vertex values. Specifically, we let $N$ be the maximum number of vertices we wish to consider at any time. At each value of $x_t$, we enumerate all assignments, discard infeasible ones, and then sort the remaining $K$ assignments by $\mathrm{sd}_k(x_t)$ value, and place $\mathrm{sd}_k(x_t)$ expressions of the $N$ smallest ones into the slots as our signed distance constraints. If $K<N$, we copy the $\mathrm{sd}_k(x_t)$ expression with the largest $sd_k(x_t)$ value to fill the remaining slots, because the larger values are less important and correspond to inactive constraints. Choosing $N$ too small can adversely affect the algorithm success rate, but larger $N$ is more expensive to compute. The upper bound for meaningful $N$ is the most complicated situation when there is vertex to vertex contact. In the 2-dimensional case, $N\leq 4$ (two constraints from the ego and obstacle each), and more for higher dimensions, scaling with the sum of the maximum number of faces forming a vertex for the ego and obstacle.

The slot concept we introduce does not completely address the issue at hand. As $x_t$ is updated in the optimization, the particular vertices which fill each slot can change. Specifically, since each vertex is an assignment point, the assignment points which fill each slot can change. Although the assignment points are themselves smooth functions of $x_t$, this implies that the output value of each slot is not a smooth function of $x_t$. Nevertheless, we propose simply ignoring this fact when computing the derivative of each constraint slot (as needed by optimization solvers), and simply treating the derivative of the assignment point value (for the given value of $x_t$) as the derivative of the constraint slot. 

The reformulation we have so far presented may seem counterintuitive. Initially, the signed-distance constraint appearing in \cref{opt} was a continuous but non-differentiable constraint. Using the slot formulation of \cref{opt2}, there are $K$ constraints which are also continuous but non-differentiable. %
However, there are two key benefits to this reformulation.
The first benefit of this new formulation is representing derivative information at points of non-differentiability. In the formulation \cref{opt2}, non-differentiability occurs whenever there are multiple assignment points which could be placed into the constraint slots due to equivalent value. The only constraint  slots which can ever be ``active'' are the first slot, and any other slots who are tied in value at a particular iteration. After some perturbation to $x_t$, the values will diverge and only the smaller of the two will remain relevant for the purposes of the optimization \cref{opt2}. It is plausible that the derivative information for that slot was incorrect before the perturbation, but if that was the case, then the correct derivative information must have been associated to one of the other tied slots. This enables the solver to essentially capture the subdifferential of the signed distance at these points of non-differentiability.

\begin{figure}[b]
\centering
\includegraphics[width=\linewidth]{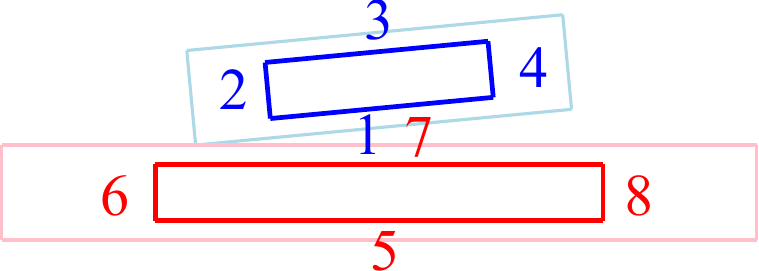}
\caption{Blue rectangle is the ego, while red rectangle is the obstacle, and ego wants to get close obstacle. Faded rectangles are the scaled ego and obstacle with $\alpha=\mathrm{sd}(x_t)$. }
\label{fig:local_info}
\end{figure}

The second benefit is related to the first one, which is the inclusion of nonlocal information. A common challenge for off-the-shelf Newton-type solvers when handling nonsmooth or nearly nonsmooth constraints is that they are completely unaware of inactive constraints which may ``suddenly'' become active.
For example, consider the configuration shown in \cref{fig:local_info}, and we want to pack these two rectangles as close as possible to each other.
If we slightly disturb to the orientation of ego, the optimal assignment may change abruptly, e.g. from assignment $(1, 2, 7)$ to assignment $(1,4,7)$, even if the signed function $\mathrm{sd}(x_t)$ remains continuous. 
If the ego is exactly parallel to the obstacle, both assignments are optimal, and $\mathrm{sd}(x_t)$ becomes nonsmooth. While this configuration is a measure zero set, it commonly occurs near the solution for packing type problems.

We find that the nonlocal information provided by the $K$ feasible assignments offers more benefit than the occasional nonsmoothness introduced by the slots or the inclusion of assignments that do not correspond to vertices. As demonstrated in the next section, our simulations and comparisons show that this method, is both simple to implement and effective in practice.

\section{SIMULATIONS}
\label{sec:simulations}
In this section we compare our problem formulation to baseline problem formulations. We use the same underlying commercial mixed complementarity problem (MCP) solver PATH Solver \cite{ferris_interfaces_1999} to solve all formulations with identical solver parameters. 
We refer to the object that we control as the ego. We assume that the obstacles are fixed, however extension to dynamic obstacles is straightforward.

\subsection{Baselines}

There are various ways to encode obstacles for optimization-based collision detection in trajectory optimization frameworks. The most common approach is to use bounding shapes.
We do not use this approach as a baseline, because we cannot directly compare our formulation for exact collision detection to unions of spheres.
The most straightforward approach for exact collision detection is encoding the KKT conditions of the low-level \cref{eq:sd_lp} directly into the high-level. The main limitation of combining the top-level and the low-level this way is that the resulting problem is no longer bilevel, and this prevents the top-level problem from reasoning about the low-level problem. We found that this approach does not work for anything other than single ego and single obstacle simple packing, and we do not include the direct KKT approach as a baseline, either. 

The first baseline method we compare to is the separating hyperplanes method \cite{nair_collision_2022,boyd_convex_2004}. It is a common and efficient approach that utilizes the hyperplane separation theorem, which states there exists a separating hyperplane that separates two nonempty nonintersecting closed polyhedral objects. It is very fast for simple settings, where there is one ego and one obstacle. 
The second baseline method is  \cite{tracy2023differentiable}, where the authors formulate the collision detection problem using a signed distance function as a convex optimization problem with conic constraints. However, this approach is flawed, because when inequality constraints are weakly active at the solution of the low-level problem, the implicit function theorem cannot be evoked, which our formulation avoids.

We compare our method to the baselines in terms of speed and reliability. Our metric for speed is the mean MCP solve time as solved by the PATH Solver. The reliability metric is more nuanced, we provide two metrics to represent reliability: (i) Success rate in terms of MCP solution achieved by PATH Solver represents the reliability of the method in successfully converging to a local optimum, (ii) and the mean cost of successes indicates how reliably the formulation achieves a low-cost local optimum. The success rate is the ratio of successful solutions to the total number of samples. 

\subsection{Benchmark Problems}

We compare our formulation and comparison baselines on the following 2-dimensional problems: (1) Simple packing problem with rectangular ego getting as close as possible to rectangular obstacle, (2) simple gap problem with rectangular ego passing through a small gap, (3) piano problem with rectangular ego navigating through L-shaped corridor, (4) random packing with rectangular ego packing into randomly generated obstacles, (5)  L through gap with L-shaped ego going through a small gap, and (6) random L packing with L-shaped ego packing into randomly generated obstacles.

\begin{figure*}[t]
\centering
\includegraphics[width=\linewidth]{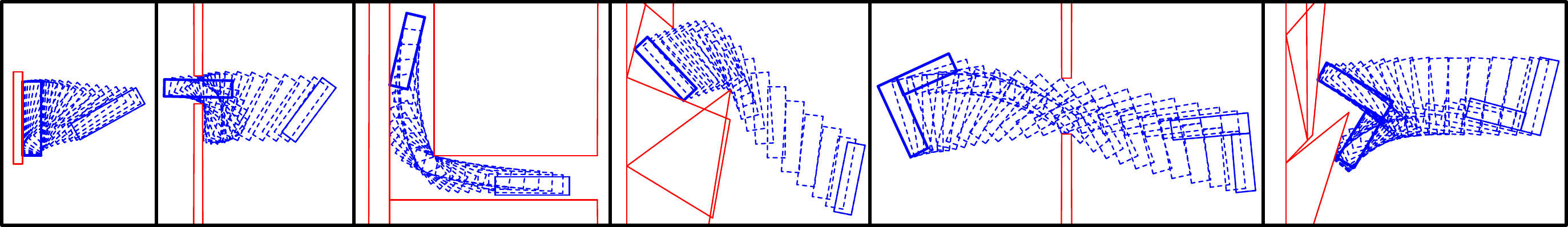}
\caption{Nonsmooth signed distance enumeration formulation examples on problems (left to right): (1) Simple packing, (2) simple gap, (3) piano, (4) random packing, (5) L through gap, (6) random L packing. The intermediate time steps shown in dashed lines, the final position shown in bold.}
\label{fig:examples}
\end{figure*}

An example solution using our method for each of these problems is shown in \cref{fig:examples}.
Throughout this section, we use identical cost function, dynamics and other simulation parameters as far as it's applicable for all baselines and problems: $T=20$ time steps, $\Delta t=0.2$ seconds of step size, quadratic control penalty factor $R=\mathrm{diag}(10^{-3},\,10^{-3},\,10^{-5})$, quadratic cost penalty factor $Q=\mathrm{diag}(2\times 10^{-3},\,2\times 10^{-3},\,0)$, $u_\mathrm{max}={[10, 10, \pi]^\intercal}$. For our method, we use $N=4$ signed distance slots. While we chose to use identical parameters for all problems for the purposes of comparisons, we note that different formulations could benefit from fine tuning of these parameters for different problems. By using the same parameters across all problems, we hope to make a fair comparison without cherry-picking favorable values. We randomly generate the obstacles to obstruct the ego's goal in random packing problems, while in piano and gap problems, the corridors and gaps are smaller than the largest dimension of the ego. %

\subsection{Results}

In this section we discuss large-scale 2-dimensional simulation results for our method and baselines on randomized obstacles and initial conditions for different problems. As we will show in this section, our method achieves higher success rates in settings where the ego needs to navigate through and pack into multiple obstacles, but it is slower than separating hyperplane formulation.
We compare our method against two baseline methods for six different problems for randomly generated initial conditions and obstacles.

\begin{table}[b]
\centering
\caption{Success rates (\%) ($n=1000$) }
	\begin{tabular}{rSSS}
	\toprule
    Problem & {Ours} & {Sep. Planes} & {\cite{tracy2023differentiable}} \\
	\midrule
    Simple packing &  \bftabnum 100 & \bftabnum 100 & 15.0 \\
	Simple gap & \bftabnum 99.9  &  89.8 & 81.8 \\
	Piano & 90.1  &  39.9 & \bftabnum 100\\
 	Random packing & \bftabnum  98.4  &  93.3 & 54.1  \\
  	L through gap & \bftabnum  99.7 &  86.8 & 32.0 \\
   	Rand. L packing & \bftabnum 99.3  & 97.1 & 39.2 \\
 	\bottomrule
	\end{tabular}
\label{tab:success_rate}
\end{table}

\begin{table}[b]
\centering
\caption{Mean successful high-level costs ($\times 10$)}  
	\begin{tabular}{
     r
     S[table-format=2.4(2)]
     S[table-format=2.4(2)]
     S[table-format=2.4(2)]
     S[table-format=table-format=1.1e1]
     }
	\toprule
    Problem & {Ours} & {Sep. Planes} & {\cite{tracy2023differentiable}}\\
	\midrule
    Simple packing &  \bftabnum 0.600(10) & 0.625(10) & 0.687(24) \\
	Simple gap &  6.215(72)  &  6.88(11) & \bftabnum 6.08(13)  \\
	Piano &  \bftabnum 4.450(24)  & 8.03(52)  & 4.466(18) \\
 	Random packing & \bftabnum  6.153(86)  &  \bftabnum 6.075(88) & 6.28(13) \\
  	L through gap &  11.485(84) & 11.635(86) & \bftabnum 10.23(23) \\
   	Rand. L packing & \bftabnum 6.525(87)  & \bftabnum 6.548(93) &  6.88(15) \\
 	\bottomrule
\end{tabular}
\label{tab:cost}
\end{table}

We showcase the success rate of our method compared to baselines in \cref{tab:success_rate}, where we use boldface to indicate the highest success rates. 
We see that the separating hyperplane approach is less reliable locally than our method for all but the simplest problem, simple packing.
Our method achieves greater than $90\%$ success for all problems, beating the baselines in all settings except for the piano problem.
We show the resulting cost for the successful solutions in \cref{tab:cost} (failed MCP solutions are not included in this metric), where we use boldface to indicate the best cost. \cref{tab:cost} tells us how successful each formulation is in terms of global success, such as being able to pass through a gap to reach the goal position. Our method outperforms the baselines in packing type problems, which means it packs more reliably compared to the baselines. In gap type problems, our method does not pass through the hole as often as \cite{tracy2023differentiable} does. We believe this is because \cite{tracy2023differentiable} tries to minimize contact with obstacles, while our method may generate a trajectory that maintains contact.

We compare the time it takes for the PATH Solver to reach a successful solution in \cref{tab:solve_time}, where we used boldface to indicate fastest computation times. Our method is slower than separating hyperplanes, however it is much more reliable, as we saw in terms of success rate (\cref{tab:success_rate}) and mean costs (\cref{tab:cost}). The separating hyperplanes method is fast and efficient for simple problems, but it struggles with complex collision detection problems where the ego and the obstacles are represented by multiple polyhedral shapes. The separating hyperplanes formulation struggles with navigating around corners, most notably in the piano problem, where the ego gets stuck at the L-shaped bend of the corridor. 
In simple gap and L through gap problems, our formulation achieves comparable mean cost (\cref{tab:cost}) to \cite{tracy2023differentiable}, at the fraction of the computation time (\cref{tab:solve_time}).
We note that \cite{tracy2023differentiable} struggles when the goal position is infeasible as it can be in the packing problems. We hypothesize this approach struggles with packing problems due to incorrect encoding of the derivatives, as previously mentioned.

\begin{table}[b]
\centering
\caption{Mean successful PATH Solver times (s)}
	\begin{tabular}{
     r
     S[table-format=2.4(2)]
     S[table-format=2.4(2)]
     S[table-format=2.4(2)]
     }
	\toprule
    Problem & {Ours} & {Sep. Planes} & {\cite{tracy2023differentiable}} \\
	\midrule
    Simple packing &   0.285(17)  & \bftabnum 0.0091(11)   & 1.23(31)  \\
	Simple gap & 0.480(30)  & \bftabnum 0.191(18) & 2.74(19)  \\
	Piano & 1.36(11)  & \bftabnum 1.124(29) & 2.515(49)  \\
 	Random packing & 1.62(17) &  \bftabnum 0.316(24) & 5.17(60)  \\
  	L through gap &  2.00(19) & \bftabnum 0.470(43) & 11.6(12)\\
   	Rand. L packing &  4.10(44)  & \bftabnum 0.407(32) &  15.6(1.2) \\
 	\bottomrule
\end{tabular}
\label{tab:solve_time}
\end{table}

Our nonsmooth signed distance enumeration method achieves superior accuracy solving complex collision detection problems, while remaining competitive in terms of performance. We have demonstrated that our formulation is more reliable compared to the separating hyperplane formulation, however it is slower. This is not surprising, as separating hyperplane formulation is quite efficient provided with a good initialization. Further, we show that our method is more reliable than a comparable but flawed approach in the literature for the problem of polygon collision detection, especially for packing problems where incorrect derivatives have a higher chance of becoming a problem for the solver.

\subsection{3-Dimensional Experiments}

In this section we present a limited comparison of our method in the 3-dimensional setting for the random packing experiment, and we consider separating hyperplanes as the only baseline, because we chose to focus on the extensibility of our method. We expect the results from the 2-dimensional experiments to fully carry on to the 3-dimensional case, however the increased complexity requires more computations, and we leave a more detailed comparison in 3-dimensions for future work.
We assume the ego and the obstacle are tetrahedrons, and we consider $T=2$ and $\Delta t=2$ seconds for simplicity. The results of our experiment is provided in \cref{tab:3d_simple_packing}. We see a trend similar to the 2-dimensional case. Our method achieves higher success rate than the separating hyperplane, however, it is significantly slower in  terms of computation time. The two approaches are comparable in terms of ability to find low-cost local optima.

\begin{table}[t]
\centering
\caption{Comparison in 3-Dimensional random packing ($n=1000$)}
	\begin{tabular}{
     r
     S[table-format=2.4(2)]
     S[table-format=2.4(2)]
     S[table-format=2.4(2)]
     }
	\toprule
    Simple Packing & {Ours} & {Sep. Planes}\\
	\midrule
 	Success rate (\%) & \bftabnum 88.9  & 70.4 \\
    Cost ($\times 10$) &  1.1254(84) &  1.176(14) \\
    Time (s) & 1.275(80) & \bftabnum 0.155(13) \\
 	\bottomrule
	\end{tabular}
\label{tab:3d_simple_packing}
\end{table}

\section{CONCLUSION}
\label{conclusion}

In this paper we proposed a procedure for dealing with nonsmooth constraints by enumerating the values obtained at all vertices of the feasible region of signed distance between polyhedra. We then demonstrated our formulation on several challenging collision detection problems in 2 and 3-dimensional settings, and compared them to common existing formulations. We found that our method outperforms common baselines in terms of success rate. Our formulation is slower, but more reliable than separating hyperplanes.  

Our method is not limited to the collision detection problem. It can be extended to other nonsmooth optimization problems, provided the source of nonsmoothness are constraints in the form of point-wise optimum functions composed of multiple smooth functions. In the future, we plan to investigate extension of nonsmooth signed distance assignment enumeration to general convex shapes.

\bibliographystyle{IEEEtran}
\bibliography{references}

\addtolength{\textheight}{-12cm}   %

\end{document}